# Anatomical Connections of Primate Mediodorsal and Motor Thalamic Nuclei with the Cortex

Bianca Sieveritz [1], Roozbeh Kiani [1,2]

[1] Center for Neural Science, New York University, New York, USA

[2] Department of Psychology, New York University, New York, USA

Corresponding authors:

Bianca Sieveritz, bs4226@nyu.edu

Roozbeh Kiani, roozbeh@nyu.edu

## Abstract

Non-sensory thalamic nuclei interact with the cortex through thalamocortical and cortico-basal ganglia-thalamocortical loops. Reciprocal connections between the mediodorsal thalamus (MD) and the prefrontal cortex are particularly important in cognition, while the reciprocal connections of the ventromedial (VM), ventral anterior (VA), and ventrolateral (VL) thalamus with the prefrontal and motor cortex are necessary for sensorimotor information processing. However, limited and often oversimplified understanding of the connectivity of the MD, VA, and VL nuclei in primates have hampered development of accurate models that explain their contribution to cognitive and sensorimotor functions. The current prevalent view suggests that the MD connects with the prefrontal cortex, while the VA and VL primarily connect with the premotor and motor cortices. However, past studies have also reported diverse connections that enable these nuclei to integrate information across a multitude of brain systems. In this review, we provide a comprehensive overview of the anatomical connectivity of the primate MD, VA, and VL with the cortex. By synthesizing recent findings, we aim to offer a valuable resource for students, newcomers to the field, and experts developing new theories or models of thalamic function. Our review highlights the complexity of these connections and underscores the need for further research to fully understand the diverse roles of these thalamic nuclei in primates.

## 1. Introduction

Cognitive functions depend on interactions between non-sensory thalamic nuclei and cortex, mediated through thalamocortical or cortio-basal ganglia-thalamocortical loops (Catanese and Jaeger, 2021; Mitchell, 2015; Takahashi et al., 2021). Earlier studies have established important roles for these nuclei in working memory, decision making, and goal-directed motor behavior. For example, the mediodorsal thalamus (MD) and its reciprocal connections with the prefrontal cortex play a role in memory acquisition (Mitchell and Gaffan, 2008; Mitchell et al., 2007a), adaptive decision-making (Browning et al., 2015), updating goal values (Alcaraz et al., 2018; Chakraborty et al., 2016), and reward devaluation (Mitchell et al., 2007b). Moreover, reciprocal connections between the motor thalamus and the anterior lateral motor cortex are necessary for sensorimotor information processing in rodents (Guo et al., 2017). These results encourage further investigation of response dynamics of thalamic neurons and the effect of their perturbations on behavior. While the frequency of such studies is increasing in the rodent brain, much-needed studies on the primate brain are lagging behind. In this review, we organize current knowledge about the anatomical connectivity of the primate MD, ventral anterior (VA), and ventrolateral (VL) thalamus in an accessible format, hoping to reduce barriers for new studies.

We are writing this review to address a clear need in the field, as recent advancements have provided new insights into thalamic connectivity that warrant a comprehensive update. By synthesizing this information, we aim to provide a resource that will be valuable to multiple audiences: students who need a concise and comprehensive overview, newcomers to the field who would benefit from a thorough and accessible introduction, and experts who seek to advance new theories or develop quantitative models of thalamic functions. Additionally, our review identifies gaps and areas for future investigation.

Previous summaries of primate MD, VA and VL connectivity with the cortex often left out weaker connections for simplification. For example, the prevalent view shaped by those summaries is that MD connects with the medial, orbital, and dorsolateral prefrontal cortex, while VA and VL connect with the premotor and motor cortex (Haber and McFarland, 2001). However, several past findings fall outside this simplified picture. For example, subregions of VL have been reported to also project to the dorsomedial and dorsolateral prefrontal cortex as well as to the frontal eye fields (Fang et al. 2006; Nakano et al. 1992). While weak connections might indicate a smaller effect on neurons in the target area, this is not always the case. Additionally, many earlier attempts to summarize connectivity of MD, VA and VL with cortex focused primarily on the prefrontal, premotor and motor cortex, neglecting that these thalamic nuclei also have diverse connections with the temporal, parietal, limbic, and insular lobes.

Here, we provide a comprehensive review of MD, VA and VL connections with the cortex in non-human primates. Understanding these connections is a crucial first step in forming hypotheses about how these thalamic nuclei might contribute to cognition. While anatomical studies do not provide information on the strength of these connections, they do offer insights into which cortical areas can – and cannot – directly exchange information with MD, VA and VL.

## 2. Anatomical subdivisions of the motor thalamus

In primates and rodents, the motor thalamus includes ventral and lateral thalamic nuclei that receive cerebellar, pallidal, and nigral inputs and have reciprocal connections with the motor cortex. In primates, a variety of subdivisions have been proposed for the motor thalamus, causing confusion in the literature and frustration for newcomers (see Percheron et al., 1996 for an extensive overview of different nomenclatures). Some of the better-known nomenclatures include those introduced by Vogt (Vogt and Vogt, 1941; Vogt, 1909), Olszewski (Olszewski, 1952), Ilinsky (Ilinsky and Kultas-Ilinsky, 1987), and Jones (Jones, 1990). The Vogts' pioneering work in old world monkeys (Vogt, 1909) and humans (Vogt and Vogt, 1941) primarily focused on cytoarchitecture, which resulted in thalamus being divided into over 40 nuclei and subnuclei (García-Cabezas et al., 2023). Hassler built on Vogts' parcellation (Hassler, 1959; Hassler et al., 1979), but many neuroscientists later pointed out that Hassler's system divides the thalamus into too many nuclei and subnuclei – for example, what is known as the VA in later nomenclatures consists of 8 subdivisions in Hassler's system (Percheron et al., 1996).

The Vogts' and Hassler's methods for dividing the motor thalamus, while pioneering for their time, were also limited in terms of linking specific thalamic regions to their functional roles. The 'Anglo-American' or 'Michigan' school of thought in thalamic research emerged in the mid-twentieth century as a response to these limitations. With the assumption that the thalamic regions connecting to the same brain areas have similar functional roles (Walker, 1938), these early approaches in the 'Anglo-American' school aimed to divide thalamus into regions with a homogenous cytoarchitecture that receive input from only one brain area or project to only one particular cortical area in a manner preserved across species (Percheron et al., 1996). This approach ultimately proved problematic for motor-related thalamic nuclei, because their anatomical connections are too diverse and their cytoarchitectural borders fuzzy and poorly matching with the boundaries defined by inputs or outputs. A prominent study illustrating this point reported that neither VA nor VL has a specific cortical projection target, but instead projects to several different cortical areas (Kievit and Kuypers, 1977).

To address these challenges, Ilinsky and Kultas-Ilinsky (1987) as well as Jones (1990) proposed new systems to divide the motor thalamus based on subcortical inputs and neurological functions. Ilinsky and Kultas-Ilinsky (1987) observed that motor thalamus comprises three zones that solely receive input from substantia nigra, globus pallidus, or deep cerebellar nuclei. Jones (1990) defined VA as the division receiving input from substantia nigra, VLa as the division receiving input from globus pallidus, and VLp as the division receiving input from cerebellar spinal nuclei. VLp also receives inputs from the spinothalamic tract (Asanuma et al., 1983b, 1983a) as well as sparse vestibulothalamic projections (Asanuma et al., 1983a; Lang et al., 1979). In contrast to Jones (1990), Ilinsky and Kultas-Ilinsky (1987) included both the nigral and pallidal zones in VA, which resulted in one coherent nucleus that primarily receives input from the output nodes of basal ganglia. They defined VL as the area receiving input from deep cerebellar nuclei. VA was further subdivided into a magnocellular (VAmc), a parvicellular (VApc), and a densicellular (VAdc) region. VAmc solely receives input from substantia nigra pars reticulata (SNr), VApc receives input from the medial internal globus pallidus (GPi), and VAdc receives input from the lateral GPi. In addition to these inputs, a recent study in macaques has found direct inputs from the subthalamic nucleus to the whole VA-VL complex (Rico et al., 2010). Some macaque nomenclatures include a VM nucleus (Jones, 1990; Walker, 1938), but most integrate it in either VA, VL, or VPM (Jones, 1985). Ilinsky and Kultas-Ilinsky (1987) were unable to identify a unique VM nucleus in macaques and included it in VA.

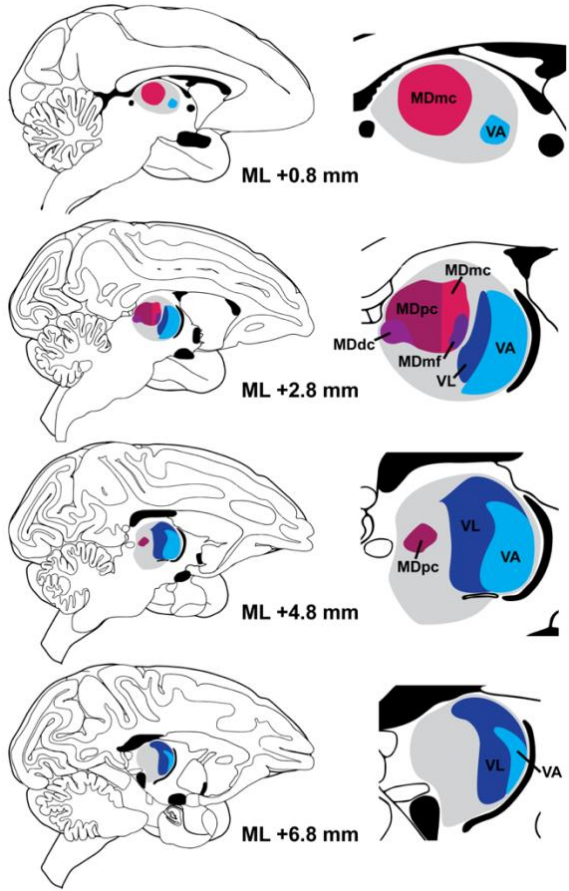

**Figure 1.** Boundaries of the thalamus in the brain (gray), and the location of MD and its subdivisions, VA, and VL according to Ilinsky and Kultas-Ilinsky (1987). The mediolateral position of each parasagittal slice is defined with respect to the inter-hemispheric midline.

In summary, the names of motor thalamic nuclei, their boundaries, and their subdivisions have undergone significant changes over time, reflecting the evolution of techniques and conceptual frameworks in neuroscience. Both Jones's and Ilinskys' nomenclature have been widely adopted by non-human primate neuroscientists, but neurosurgeons still use an earlier nomenclature rooted in Hassler's subdivisions (García-Cabezas et al., 2023; Percheron et al., 1996). In this review, we will use Ilinsky's and Kultas-Ilinsky's (1987) subdivisions of the motor thalamus (VA and VL), and mediodorsal thalamus (Fig. 1). Given that Ilinsky and Kultas-Ilinsky (1987) were unable to identify a unique VM nucleus in the macaque, we will treat VM as a part of VA in this review. Table 1 maps Ilinsky and Kultas-Ilinsky's motor thalamic subdivisions to those from other commonly used nomenclatures.

| Thalamic nucleus in this review | Ilinsky & Kutas-Ilinsky, 1987 | Jones (1985) | Olszewski (1952) | Hassler (1959) | Vogt (1941) | Walker (1938) | Subcortical afferents |
|---|---|---|---|---|---|---|---|
| VA | VAmc | VAmc | VAmc | Lpo.mc | - | - | SNr |
|  | VApc | VApc | VApc | Lpo | - | VA | Medial part of GPi |
|  | VAdc | VLa | VLo | Voa | Voe | VL | Lateral part of GPi |
|  | VM | VM | VLm | Vom Voi | Voi | VM | No distinct subcortical input |
| VL | VL | VLp | VLc VLps VPLo Area X | Vop Vim | Vim | VL | Deep cerebellar nuclei, spinal cord, vestibular nuclei |

**Table 1.** Mapping of motor thalamic nuclei across nomenclatures according to Ilinsky and Kultas-Ilinsky (1987; adapted from table 2 in Ilinsky and Kultas-Ilinsky, 1987). SNr: substantia nigra pars reticulata; GPi: internal globus pallidus.

## 2.1. Motor thalamic nuclei in rodents

Similar to non-human primates, motor thalamus in rodents consist of VA, VL, and VM (Bosch-Bouju et al., 2013; Kuramoto et al., 2011; Sieveritz et al., 2019), even though it is unclear whether they map onto their primate namesakes. Specifically, rostroventral VA and VM primarily receive palladio-nigral inputs, while caudodorsal VA and VL primarily receive cerebellar inputs (Kuramoto et al., 2011). Palladio-nigral axon terminals are primarily gabaergic, while cerebellar axon terminals are primarily glutamatergic (Kuramoto et al., 2011). Immunohistochemistry for gabaergic and glutamatergic neurotransmitters is commonly used to discriminate between the parts of the motor thalamus that receive palladio-nigral and cerebellar inputs. It is unclear whether the rodent VA and VL are exact homologues of the primate VA and VL. If mapped according to their inputs, rostroventral VA and VM roughly map onto the primate VA, and caudodorsal VA and VL roughly map onto the primate VL.

## 3. Anatomical subdivisions of the mediodorsal thalamus

The definition of MD and its subdivisions have been largely consistent in past studies. Most nomenclatures divide MD based on cytoarchitecture into a medial magnocellular region (MDmc), a central parvocellular region (MDpc), and a lateral region (Mitchell, 2015). Many nomenclatures further divide the lateral MD into a caudal densocellular region (MDdc; Bachevalier et al., 1997; Bentivoglio et al., 1993; Jones, 1985), and a rostral part known as pars multiforms (MDmf). MDmc is magnocellular in Nissl preparations. It has large, evenly spaced cells that are embedded in dense, fibrous neuropil (Jones, 1985). In contrast, MDpc has less fibrous neuropil and smaller cells, which are more variable in size (Jones, 1985). The lateral MD stains more heavily for acetylcholinesterase (Cavada et al., 1995), and consists of large cells (Jones, 1985).

MDmc has two subdivisions with distinct cytoarchitecture. The lateral part of MDmc is defined by fine myelinated fibers and has been named the MDmc pars fibrosa. In contrast, the medial part of MDmc is defined by poor myelination and has been named the MDmc pars paramediana (Ray and Price, 1993). Most studies do not discriminate between the two MDmc subdivisions and we will follow this precedent.

Some neuroanatomical studies do not distinguish between MD subdivisions, even though cytoarchitecture and neuroanatomical connections differ between them. In this review, we primarily rely on the studies that distinguish MDmc, MDpc, MDdc and MDmf. However, when appropriate, we also included studies that do not distinguish different MD subdivisions.

### 3.1. Subcortical inputs to the mediodorsal thalamic nucleus

In primates, subcortical inputs to MD are more diverse than those to the VA-VL complex. MDmc receives subcortical inputs from rostromedial SNr (Carpenter et al., 1976; Ilinsky et al., 1985), rostrolateral GPi (Carpenter et al., 1976; Ilinsky et al., 1985), ventral pallidum (Russchen et al., 1987), the subiculum (Saunders et al., 2005), substantia innominata (Russchen et al., 1987), and the amygdala (Aggleton and Mishkin, 1984; Russchen et al., 1987). In contrast, MDpc receives subcortical inputs only from the lateral SNr (Ilinsky et al., 1985; Tanibuchi et al., 2009). The lateral MD (MDdc and MDmf) receives subcortical inputs from the claustrum, superior colliculus, and the ventral midbrain (Erickson et al., 2004). MDmf additionally receives inputs from the dorsal part of rostromedial SNr (Middleton and Strick, 2000). MD further projects to and receives inputs from all parts of the reticular thalamus (Kuroda and Price, 1991).

### 4. Connections of MD, VA, and VL with cortex

MD, VA, and VL have connections with many cortical areas; some of which are reciprocal. Summarizing these connections across studies requires a shared system for identifying cortical regions. However, just like thalamic subdivisions which vary across published literature, naming conventions for cortical areas are diverse across studies. To overcome this challenge, we mapped cortical areas in the cited studies onto Brodmann areas using descriptions of naming conventions within each study, figures illustrating the position of specific cortical areas, and stereotaxic coordinates. Broadman parcellation of the prefrontal cortex was refined by both Walker (1940) and Carmichael and Price (1994). The refined Broadman parcellation of the prefrontal cortex by Carmichael and Price (1994) is commonly used today and we have adapted it here. To help readers orient, Figure 2 illustrates parcellations of the macaque cortex in the Brodmann system.

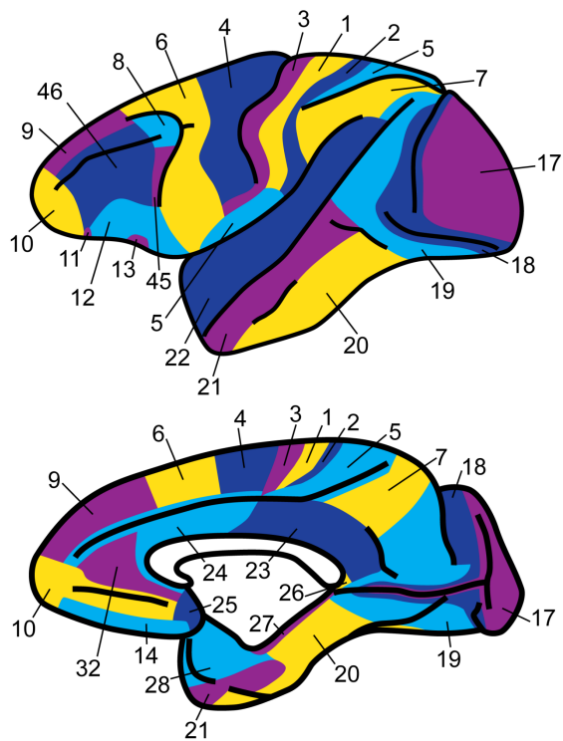

**Figure 2.** Brodman areas of the macaque cortex.

## 4.1. Prevalent theories on the organization of primate thalamocortical connections

Parallel loop models propose that cortical, basal ganglia and thalamic areas that are associated with similar functions form anatomically segregated loops (Alexander et al., 1986; Middleton and Strick, 2000). For example, Middleton and Strick (2000) proposed that visual information is processed in one loop formed between the inferior temporal area TE, striatum, substantia nigra pars reticulata (SNr), and VA; cognitive information is processed in a second loop formed between the prefrontal cortex, striatum, the globus pallidus internal (GPi) and SNr, and MD/VA; oculomotor information is processed in a third loop formed between frontal eye fields, striatum, SNr, and MD/VA; and motor information is processed in a fourth loop formed between motor cortical areas, striatum, the GPi, and VL. In this model, thalamic neuronal populations participating in each loop do not overlap, and information exchange between the loops happens through interactions between cortical areas.

On the opposite extreme of frameworks conceptualizing how information flows between cortical areas, the basal ganglia, and thalamus would be a model which proposes that thalamic nuclei have reciprocal connections with a multitude of cortical areas, which would enable convergence of information from different functional cortical domains onto individual thalamic nuclei and divergence from those nuclei to a multitude of cortical domains.

Haber and McFarland (2001) proposed a model situated in between these two extremes that proposes three cortico-basal ganglia-thalamocortical loops: a limbic loop formed by the medial, orbital and dorsolateral prefrontal cortex with MD, a cognitive loop formed by rostral motor cortical areas with VA, and a motor loop formed by caudal motor cortical areas with VL (Fig. 3). However, in contrast to parallel loop models, this model proposes that connections between MD, cortical areas, and VA allows for bidirectional information to flow between the limbic and cognitive loops, and connections between MD, VA, cortical areas, and VL permit unidirectional information flow from the limbic and cognitive loops to the motor loop. Specifically, the medial, orbital, and dorsolateral prefrontal cortex have reciprocal connections with MD; the medial and orbital prefrontal cortex projects to VA; and the dorsolateral prefrontal and rostral motor cortex have reciprocal connections with VA. Thus, in this model, MD and VA are able to exchange information through the dorsolateral prefrontal cortex as an intermediary, allowing for bidirectional information flow between the limbic and cognitive loops. The dorsolateral prefrontal cortex and rostral motor cortex project to VL, allowing information to flow unilaterally from the limbic and cognitive loops to the motor loop. Although the model does not allow for bidirectional information flow between all the loops, it accounts for some convergence of information across functional domains.

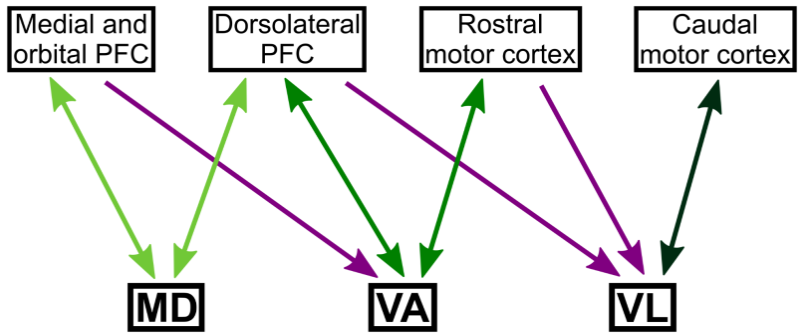

**Figure 3**. Network of connections between MD, VA, VL, prefrontal cortical and motor cortical areas as suggested by Haber and McFarland (Haber and McFarland, 2001). Green: reciprocal connections; purple: unidirectional connections from the cortex to the thalamus. PFC: prefrontal cortex.

We will discuss whether Haber and McFarland's model (Haber and McFarland, 2001) holds up, when weaker connections between the thalamus and cortex are taken into account. Haber and McFarland (2001) differentiated between the orbital and medial prefrontal cortex, the dorsolateral prefrontal cortex, the rostral motor cortex, and the caudal motor cortex. Haber and McFarland (2001) counted the frontal pole (BA 10), which is part of the polar prefrontal cortex toward the dorsolateral prefrontal cortex. Here, we will count Brodmann areas (BA) 11, 12, 13, 14, 24, 25, and 32 toward the orbital and medial prefrontal cortex, BA 9, 46 and 10 toward the dorsolateral and polar prefrontal cortex, BA 8 and the rostral part of BA 6 toward the rostral motor cortex, and BA 4 and the caudal part of BA 6 toward the caudal motor cortex.

### 4.2 MD connections with the cortex

### 4.2.1 MD connections with the prefrontal and motor cortex

Reciprocal connections of MD with the medial, orbital, polar, and dorsolateral prefrontal cortex are well established. Further, in contrast to Haber and McFarland's model (Haber and McFarland, 2001), several studies have reported that MD also has reciprocal connections with the rostral and caudal motor cortex.

Orbitofrontal cortex: MDmc and MDpc project to BA 11, BA 12 and BA 13 (Bachevalier et al., 1997; Erickson and Lewis, 2004; Giguere and Goldman-Rakic, 1988; Goldman-Rakic and Porrino, 1985; McFarland and Haber, 2002; Middleton and Strick, 2000; Ray and Price, 1993). MDmf projects only to BA 12 (Goldman-Rakic and Porrino, 1985; Middleton and Strick, 2000).The ventral part of MDmc projects primarily to BA 11 and BA 12, while the dorsal part primarily projects to BA 13 (Giguere and Goldman-Rakic, 1988). Projections from MDmc to BA 11, BA12 and BA 13 are reciprocated (Giguere and Goldman-Rakic, 1988; Ray and Price, 1993).

Medial prefrontal cortex: MDmc, MDpc, and MDdc project to BA 14 (Bachevalier et al., 1997; Giguere and Goldman-Rakic, 1988; Ray and Price, 1993). Projections from MDmc (Giguere and Goldman-Rakic, 1988) and MDpc to BA 14 are reciprocated (Ray and Price, 1993). All subdivision of MD project to BA 24 and, with the exception of MDmf, to BA 32 (Bachevalier et al., 1997; Erickson and Lewis, 2004; Giguere and Goldman-Rakic, 1988; Goldman-Rakic and Porrino, 1985; Ray and Price, 1993). Projections from all subdivisions of MD to BA 24 (Giguere and Goldman-Rakic, 1988; Ray and Price, 1993) as well as projections from MDpc to BA 32 are reciprocated (Ray and Price, 1993). MDmc additionally projects to BA 25 (Bachevalier et al., 1997), and MDpc receives axons from BA 25 (Chiba et al., 2001).

Dorsolateral and polar prefrontal cortex: MDpc and MDdc project to medial BA 9 (Erickson and Lewis, 2004; Giguere and Goldman-Rakic, 1988; Goldman-Rakic and Porrino, 1985), and MDmc, MDpc, and MDmf project to lateral BA 9 / BA 46 (Bachevalier et al. 1997; Erickson and Lewis 2004; Giguere and Goldman-Rakic 1988; Goldman-Rakic and Porrino 1985; Middleton and Strick 2000; Tanibuchi et al. 2009). Projections from MDpc to medial BA 9 and lateral BA 9 / BA 46 are reciprocated (Giguere and Goldman-Rakic, 1988). Furthermore, MDmc (Bachevalier et al., 1997; Burman et al., 2011; Goldman-Rakic and Porrino, 1985; Petrides and Pandya, 2007), MDpc (Bachevalier et al., 1997; Erickson and Lewis, 2004; Goldman-Rakic and Porrino, 1985) and MDdc (Bachevalier et al., 1997) project to the frontal pole (BA 10). Projections from MDmc to the frontal pole (BA 10) are reciprocated (Burman et al., 2011; Petrides and Pandya, 2007; Ray and Price, 1993).

Rostral motor cortex (BA8 and rostral BA6): The Haber and McFarland model (Haber and McFarland, 2001) suggests that there are no connections between MD and motor cortical areas, but several studies have reported these connections. All subdivisions of MD project to the frontal eye fields (part of BA 8; Bachevalier et al., 1997; Erickson and Lewis, 2004; Fang et al., 2006; Goldman-Rakic and Porrino, 1985; McFarland and Haber, 2002; Middleton and Strick, 2000). Projections from MDmf to the frontal eye fields are reciprocated (Giguere and Goldman-Rakic, 1988). Furthermore, MDpc projects to the rostral premotor cortex (BA 6; Erickson and Lewis, 2004; Matelli and Luppino, 1996; Rouiller et al., 1999), and MDpc (Erickson and Lewis, 2004) and MDmf project to the pre-supplementary motor area (pre-SMA; BA 6;

Wiesendanger and Wiesendanger, 1985). The pre-SMA (BA 6) projects back to MDmf (Wiesendanger and Wiesendanger, 1985), and the rostral premotor cortex (BA 6) projects back to MDpc but also sends axons to MDmf (Fang et al., 2006).

Caudal motor cortex (BA4 and caudal BA6): MDpc (Goldman-Rakic and Porrino, 1985; Wiesendanger and Wiesendanger, 1985), MDdc (Giguere and Goldman-Rakic, 1988; Goldman-Rakic and Porrino, 1985), and MDmf (Giguere and Goldman-Rakic, 1988; Goldman-Rakic and Porrino, 1985; Wiesendanger and Wiesendanger, 1985) have reciprocal connections with the supplementary motor area (SMA; BA 6). In addition, MDpc projects to the caudal premotor cortex (BA 6; Erickson and Lewis, 2004), and MDmf has reciprocal connections with the primary motor cortex (BA 4; Wiesendanger and Wiesendanger, 1985).

To conclude, MD has reciprocal connections with the medial, orbital, dorsolateral and polar prefrontal cortex, as well as the rostral and caudal motor cortex (Fig. 4). In theory, MD is able to exchange information directly with the motor cortex, bypassing VA and VL. These connections form gradients along the medial-lateral axis (Fig. 5). MDmc, the most medial part of MD, has reciprocal connections with the medial, orbital, dorsolateral, and polar prefrontal cortex, projects to the rostral motor cortex, and has no connections with the caudal motor cortex. MDpc has reciprocal connections with the medial and dorsolateral prefrontal cortex, as well as the rostral and caudal motor cortex, and projects to the orbital and polar prefrontal cortex. MDdc has reciprocal connections with the medial prefrontal as well as the caudal motor cortex, and projects to the dorsolateral prefrontal, the polar prefrontal and the rostral motor cortex. MDmf, the most lateral part of MD, has reciprocal connections with the medial prefrontal, rostral motor and caudal motor cortex. It is the only MD division with reciprocal connections to the primary motor cortex. Furthermore, MDmf projects to the orbital, and dorsolateral prefrontal cortex. Overall, the connections between different subdivisions of MD with the prefrontal and motor cortex form a systematic gradient, where the medial regions of MD connect more strongly with the medial and orbital prefrontal areas, central regions connect more strongly with the dorsolateral and polar prefrontal areas, and lateral regions with the motor cortical areas.

### 4.2.2 MD connections with non-frontal cortical areas

In addition to connections with the prefrontal and motor cortex, MD also forms connections with the parietal, temporal, limbic, and insular lobes.

Parietal lobes: parts of MD project to BA7 in the posterior parietal cortex (Gharbawie et al., 2010). Specifically, MDmc has reciprocal connections with the whole inferior parietal lobule area in BA7 (Giguere and Goldman-Rakic, 1988), which extends across areas PF (area 7b), PFG (area 7b), and PG (area 7a). MDpc and MDmf project to areas PF (area 7b), PFG (area 7b), PG (area 7a), and Opt (area 7a; Schmahmann and Pandya 1990). MDpc further projects to area PGm (area 7m; Buckwalter et al. 2008). MDdc projects to areas PG (area 7a), Opt (area 7a), and PGm (area 7m; Buckwalter et al., 2008; Schmahmann and Pandya, 1990),.

MD further projects to the superior parietal lobule (BA5). Specifically, MDpc projects to areas PEa (Schmahmann and Pandya, 1990) and PEci (Buckwalter et al., 2008) within BA 5, MDdc projects to areas PE (Schmahmann and Pandya, 1990), PEc (Buckwalter et al., 2008) and PEci (Buckwalter et al., 2008), and MDmf projects to areas PEa (Schmahmann and Pandya, 1990) and PEc (Schmahmann and Pandya, 1990). To our knowledge, projections from MDmc to BA 5 have not been reported and none of these projections are reciprocal.

Temporal lobe: MDmc forms reciprocal connections with the superior temporal gyrus (BA 22; Giguere and Goldman-Rakic, 1988; Russchen et al., 1987) and the rhinal cortex (BA 28; Bachevalier et al., 1997; Ray and Price, 1993; Russchen et al., 1987; Saunders et al., 2005), projects to area TE (BA 20; Bachevalier et al., 1997), and receives axons from the piriform cortex (BA27; Ray and Price, 1993). In addition, a study that did not differentiate between subdivisions of MD reported projections from MD to the auditory cortex (BA 22; Cappe et al., 2009).

Limbic and insular lobe: MDpc as well as MDdc project to the posterior cingulate cortex (BA 23; Buckwalter et al., 2008). MDmc has reciprocal connections with the insular cortex (Giguere and Goldman-Rakic, 1988), and receives axons from the primary olfactory cortex (Russchen et al., 1987).

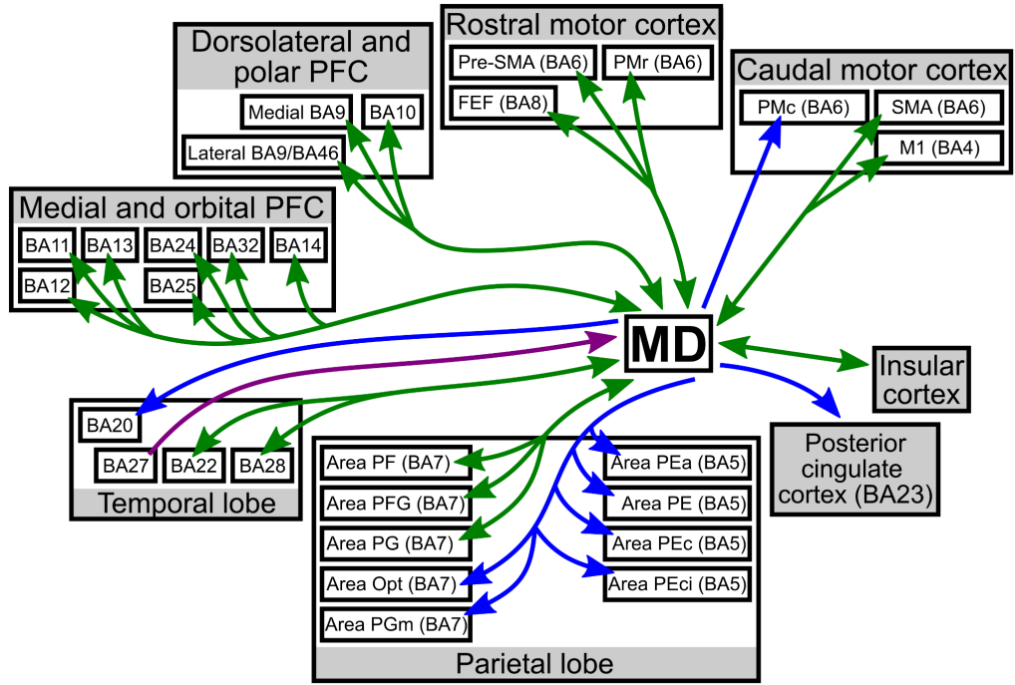

**Figure 4.** Connectivity between MD and the cortex. Green: reciprocal connections; blue: unidirectional connections from the thalamus to the cortex; purple: unidirectional connections from the cortex to the thalamus.. PFC: prefrontal cortex; FEF: frontal eye fields; PMr: rostral premotor cortex; PMc: caudal premotor cortex; SMA: supplementary motor area; M1: primary motor cortex.

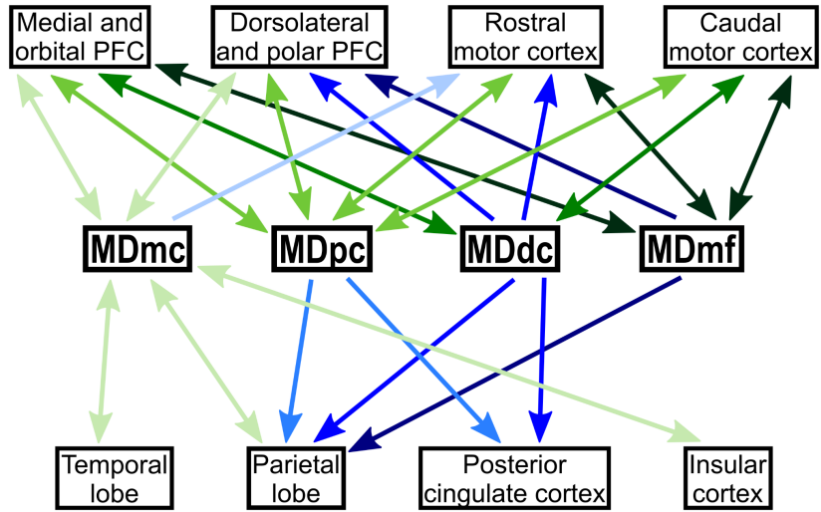

**Figure 5.** Simplified overview of the connectivity between different parts of MD and cortex. Green: reciprocal connections; blue: unidirectional connections from the thalamus to the cortex.

**4.3 VA and VL connections with cortex**

**4.3.1 VA connections with the prefrontal and motor cortex**

Unidirectional connections from the orbital and medial prefrontal cortex to VA, and reciprocal connections of the dorsolateral prefrontal and rostral motor cortex with VA are well established. Further, in contrast to Haber and McFarland's model (Haber and McFarland, 2001), several studies have reported that connections between the orbital and medial prefrontal cortex and VA are reciprocal, and that VA has reciprocal connections with the caudal motor cortex.

Orbitofrontal and medial prefrontal cortex: VA has reciprocal connections with BA 12, BA 13, and BA 24 (Goldman-Rakic and Porrino 1985; McFarland and Haber 2002; Middleton and Strick 2000; Nakano et al. 1992; Tanibuchi et al. 2009; Xiao et al. 2009), and receives projections from BA 32 (Chiba et al., 2001; Xiao et al., 2009).

Dorsolateral and polar prefrontal cortex: VA has reciprocal connections with medial BA 9 (Goldman-Rakic and Porrino, 1985; McFarland and Haber, 2002; Middleton and Strick, 2000; Nakano et al., 1992; Xiao et al., 2009), lateral BA 9 / BA 46 (Fang et al., 2006; Goldman-Rakic and Porrino, 1985; McFarland and Haber, 2002; Middleton and Strick, 2000; Nakano et al., 1992; Tanibuchi et al., 2009; Xiao et al., 2009), and the frontal pole (BA 10; Goldman-Rakic and Porrino, 1985; Xiao et al., 2009).

Rostral motor cortex (BA 8 and rostral BA 6): VA has reciprocal connections with the frontal eye fields (BA 8; Fang et al., 2006; McFarland and Haber, 2002; Middleton and Strick, 2000; Nakano et al., 1992; Xiao et al., 2009), pre-SMA (BA 6; Matelli and Luppino, 1996; McFarland and Haber, 2002; Wiesendanger and Wiesendanger, 1985), and rostral premotor cortex (BA 6; Fang et al., 2006; Matelli and Luppino, 1996; McFarland and Haber, 2002; Middleton and Strick, 2000; Nakano et al., 1992; Rouiller et al., 1999).

Caudal motor cortex (caudal BA 6 and BA 4): VA sends projections to the caudal premotor cortex (BA 6; Fang et al., 2006; Matelli and Luppino, 1996; McFarland and Haber, 2002; Middleton and Strick, 2000; Nakano et al., 1992; Rouiller et al., 1999), SMA (BA 6; Fang et al., 2006; Matelli and Luppino, 1996; McFarland and Haber, 2002; Nakano et al., 1992; Rouiller et al., 1999; Tanibuchi et al., 2009; Wiesendanger and Wiesendanger, 1985) and primary motor cortex (BA 4; Fang et al. 2006; Gharbawie et al. 2010; Matelli and Luppino 1996; McFarland and Haber 2002; Middleton and Strick 2000; Nakano et al.

1992; Rouiller et al. 1999; Tanibuchi et al. 2009; Wiesendanger and Wiesendanger 1985). Projections to SMA (BA 6) and primary motor cortex (BA 4) are reciprocal (Wiesendanger and Wiesendanger, 1985).

In summary, VA has reciprocal connections with the medial, orbital, dorsolateral, and polar prefrontal cortex, as well as with the rostral and caudal motor cortex (Fig. 6).This connectivity pattern enables VA to exchange information directly with all of these cortical areas without intermediary relays through MD or VL.

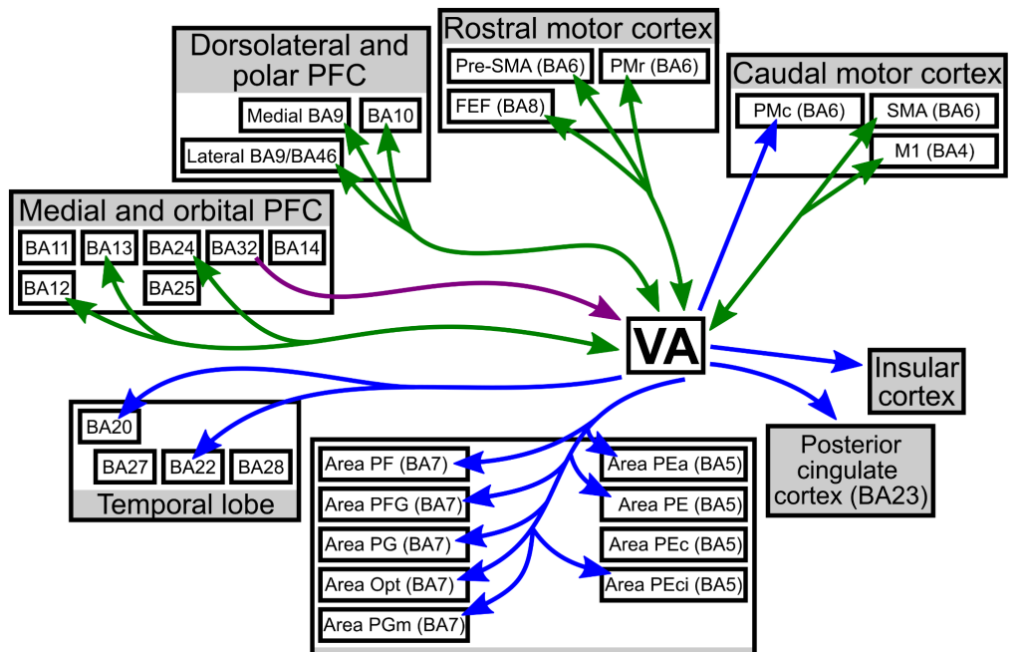

**Figure 6.** Connectivity between VA and cortex. Green: reciprocal connections; blue: unidirectional connections from the thalamus to the cortex; purple: unidirectional connections from the cortex to the thalamus. PFC: prefrontal cortex; FEF: frontal eye fields; PMr: rostral premotor cortex; PMc: caudal premotor cortex; SMA: supplementary motor area; M1: primary motor cortex.

### 4.3.2. VL connections with the prefrontal and motor cortex

Unidirectional connections from the rostral motor cortex to VL, and reciprocal connections between the caudal motor cortex and VL are well established. Further, in contrast to Haber and McFarland's model (Haber and McFarland, 2001), several studies have reported that VL has reciprocal connections with the rostral motor and dorsolateral prefrontal cortex, and receives inputs from the medial and orbital prefrontal cortex.

Orbitofrontal, medial, dorsolateral, and polar prefrontal cortex: VL has reciprocal connections with lateral BA 9 / BA 46 (Fang et al., 2006; Xiao et al., 2009). In addition, VL receives axons from medial BA 9, the frontal pole (BA 10), BA12, BA13, BA24, and BA 32 (Xiao et al., 2009).

Rostral motor cortex (BA 8 and rostral BA 6): VL has reciprocal connections with the frontal eye fields (BA 8; Fang et al. 2006; Nakano et al. 1992; Xiao et al. 2009), pre-SMA (BA 6; Matelli and Luppino 1996; Wiesendanger and Wiesendanger 1985), and rostral premotor cortex (BA 6; Fang et al. 2006; Matelli and Luppino 1996; Rouiller et al. 1999).

Caudal motor cortex (caudal BA 6 and BA 4): VL projects to the caudal premotor cortex (BA 6; Fang et al. 2006; Matelli and Luppino 1996; Nakano et al. 1992; Rouiller et al. 1999), and has reciprocal connections with both the SMA (BA 6; Fang et al. 2006; Matelli and Luppino 1996; Nakano et al. 1992; Rouiller et al. 1999; Wiesendanger and Wiesendanger 1985) as well as the primary motor cortex (BA 4; Fang et al. 2006; Gharbawie et al. 2010; Matelli and Luppino 1996; Nakano et al. 1992; Rouiller et al. 1999; Wiesendanger and Wiesendanger 1985).

In conclusion, VL maintains reciprocal connections with the dorsolateral prefrontal cortex, rostral motor cortex, and caudal motor cortex (Fig. 7). In addition, VL receives inputs from the medial, orbital, and polar prefrontal cortex (Fig. 7). This connectivity pattern enables VL to exchange information directly with the prefrontal and rostral motor cortical areas without intermediary relays through MD or VA.

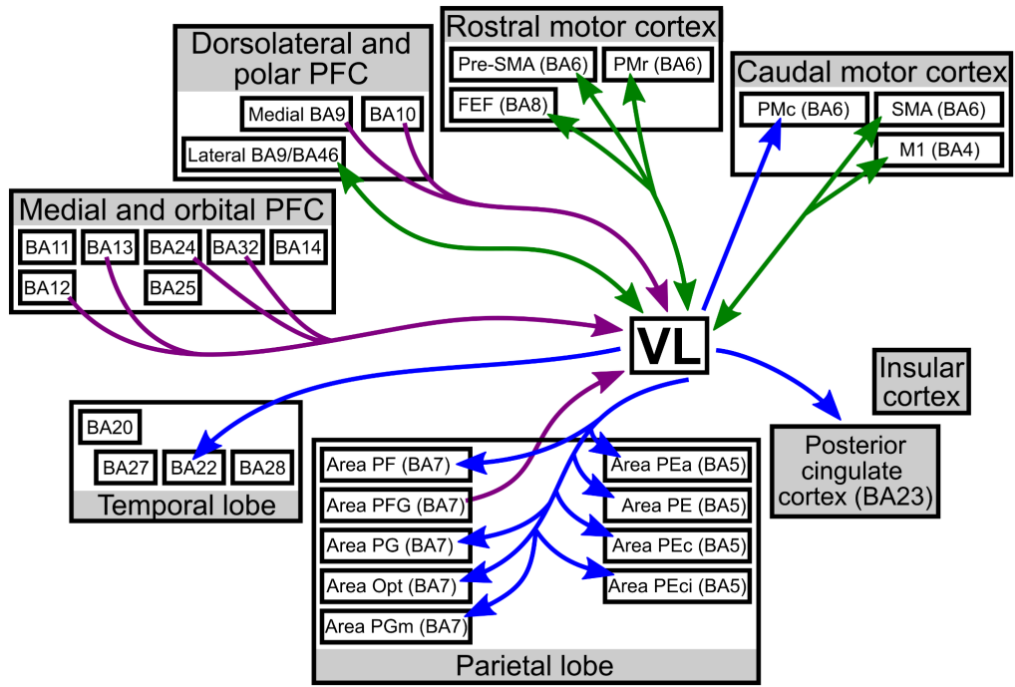

**Figure 7.** Connectivity between VL and the cortex. Green: reciprocal connections; blue: unidirectional connections from the thalamus to the cortex; purple: unidirectional connections from the cortex to the thalamus. PFC: prefrontal cortex; FEF: frontal eye fields; SMA: supplementary motor area; PMr: rostral premotor cortex; PMc: caudal premotor cortex; M1: primary motor cortex.

### 4.3.3 VA and VL connections with non-frontal cortical areas

In addition to connections with the prefrontal and motor cortex, VA and VL also form connections with the parietal, insular, temporal, and limbic, and insular lobes.

<u>Parietal lobe:</u> VA and VL both project to BA 7 in the posterior parietal cortex (Gharbawie et al., 2010). It remains unclear which specific parts of BA 7 receive input from VA, but VL specifically projects to areas PF (area 7b), PG (area 7a), Opt (area 7a) and PGm (area 7m; Buckwalter et al., 2008; Schmahmann and Pandya, 1990). To our knowledge, none of these projections are reciprocal, even though (Taktakishvili et al., 2002) reported that VL receives inputs from PFG (area 7b).

VA and VL further project to the superior parietal lobule (BA5). Specifically, both VA and VL project to areas PEa (Cappe et al., 2009; Schmahmann and Pandya, 1990), PE (Cappe et al., 2009; Schmahmann and Pandya, 1990) and PEci (Buckwalter et al., 2008). In addition, VL also projects to area PEc (Buckwalter et al., 2008; Schmahmann and Pandya, 1990).

<u>Temporal lobe:</u> VA projects to area TE (BA 20; Middleton and Strick, 2000), and VA as well as VL project to the auditory cortex (BA 22; Cappe et al., 2009).

<u>Insular and limbic lobes:</u> Both VA and VL project to the posterior cingulate cortex (BA 23; Buckwalter et al., 2008), and VA also projects to the insular cortex (Nakano et al., 1992).

## 4.4 Summary of MD, VA and VL connections with cortex

We provide a summary of MD, VA and VL connections with the cortex in Figure 8. Our summary shows that MD, VA and VL project to and receive inputs from more cortical areas than classic models acknowledge. In fact, MD, VA and VL form a large network of unidirectional and reciprocal connections with the cortex. This network allows information exchange between these thalamic nuclei and many cortical areas. For example, within the frontal lobe, MD and VA are able to directly exchange information with the medial, orbital, polar, and dorsolateral prefrontal cortex, as well as with the rostral and caudal motor cortex. VL is able to directly exchange information with the dorsolateral prefrontal, rostral motor and caudal motor cortex. In addition, VL receives inputs from the medial, orbital, and polar prefrontal cortex.

Closed-loop models as well as models that propose unidirectional flow of information from the prefrontal to motor cortex and from MD to VL do not account for these neuroanatomical findings. In contrast to the predictions made by these models, MD is able to directly exchange information with parts of the motor cortex, and VL is able to directly exchange information with the dorsolateral prefrontal cortex. However, there is a gradient in connectivity (Figure 8). MD and VA have sparser connections with the primary motor cortex and VL receives inputs from, but has no reciprocal connections with the medial, orbital, and polar prefrontal cortex. When we take into account the previously discussed gradient in connectivity within MD, the medial-anterior parts of the MD-VA-VL complex are more heavily connected with the prefrontal cortical areas, while the lateral-posterior parts are more heavily connected with the motor cortical areas.

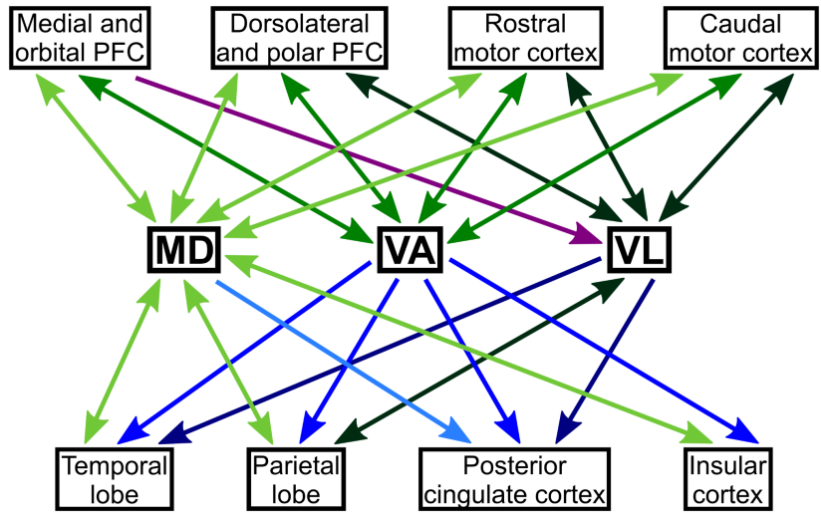

**Figure 8.** Overview of the connectivity between MD, VA, VL and cortex. Green: reciprocal connections; blue: unidirectional connections from the thalamus to the cortex; purple: unidirectional connections from the cortex to the thalamus. Connections of different thalamic nuclei have distinct color saturations. At the resolution of this figure, connections between VL and the parietal lobe are illustrated as reciprocal, but VL projects to and receives input from distinct subregions within the parietal lobe.

## 5. 'Driver' versus 'modulatory' corticothalamic projections

A prominent framework to differentiate two types of corticothalamic projections is the 'driver' and 'modulatory' framework (Sherman and Guillery, 2006). Driver inputs from the cortex to thalamus send "the main information to be conveyed to [other cortical areas]" to the thalamus (Sherman and Guillery, 2006). In contrast, modulatory inputs from the cortex to the thalamus gate the flow of information and regulate which main cortical inputs are transmitted onward to other brain areas. In this framework, driver inputs from the cortex to thalamus primarily originate in layer V, and modulatory inputs primarily originate in layer VI (Sherman and Guillery, 1998). Driver inputs preferentially innervate proximal dendrites, have thicker axons and larger terminals than modulatory inputs, and induce a larger initial excitatory postsynaptic potential. As a result, driver inputs could induce an all-or-none activation, which enables them to effectively transmit information between the cortex and thalamus, whereas modulatory inputs induce smaller initial excitatory potentials, which translates into a graded activation to control the flow of incoming information (Sherman and Guillery, 2011).

A key presumption of the driver and modulatory framework is that thalamus itself does not conduct any computation on the incoming information. Instead, thalamus acts solely as a gatekeeper and regulates which incoming information is transmitted onward. Although studies in both the visual (Saalmann and Kastner, 2011) and central thalamus (Matsuyama and Tanaka, 2021) have indicated that this might not be the case, it is still important to understand whether corticothalamic projections to MD, VA and VL originate in cortical layer V or VI. As outlined above, corticothalamic projections originating in layers V and VI induce different responses in postsynaptic neurons. These differences must be taken into account, if we want to understand the role of corticothalamic connections to MD, VA or VL in the computations that underlie cognition.

The prefrontal neurons projecting to MD are primarily located in the superficial strata of layer VI (Giguere and Goldman-Rakic, 1988; Xiao et al., 2009), with the exception of the anterior cingulate cortex, where neurons projecting to MD are equally distributed between the superficial and deep strata of layer VI (Giguere and Goldman-Rakic, 1988). Projections from prefrontal layer V to MD are scant. In contrast, in the supplementary motor neurons projecting to MD are distributed between the superficial strata of layers V and VI (Giguere and Goldman-Rakic, 1988). Similarly, in many medial, orbital and dorsolateral prefrontal cortical areas, an almost equal number of neurons projecting to VA/VL are located in layers V and VI (Xiao et al., 2009). Notable exceptions are BA 12 and BA 13, which are part of the orbitofrontal cortex, BA 46, which is part of the dorsolateral prefrontal cortex, and the ventral part of BA 8 (Xiao et al., 2009), where the majority of neurons projecting to VA/VL are located in layer VI (Xiao et al., 2009)

In summary, most corticothalamic projections from the prefrontal cortical areas to MD originate in layer VI, while corticothalamic projections from the supplementary motor cortex to MD originate in both layers V and VI. Therefore, inputs from the prefrontal cortical areas to MD might be primarily modulatory and inputs from the supplementary motor cortex to MD might be primarily driver inputs. It remains unclear how this neuroanatomical distinction contributes to differences in function. An attractive speculation is that motor cortical inputs to MD may contribute to MD sustaining and reinforcing activity of medial prefrontal neurons as reported in rodents (Francoeur et al., 2019; Miller et al., 2017).

In the case of VA/VL, we speculate that prefrontal cortical inputs to VA/VL might gate cortical motor inputs, e.g., to suppress unwanted actions, while allowing planned actions to proceed. In fact, perturbation of the cortico-thalamo-cortical loop between the VA/VL complex and the anterior lateral part of the motor cortex biases choices on a sensory discrimination task (Guo et al., 2017).

## 6. Terminal fields of thalamocortical projections

The distribution of thalamocortical terminal fields in the cortex differs significantly between the thalamocortical projections originating in MD, VA, or VL. In non-human primates, projections from MD to the prefrontal cortical areas appear to exclusively terminate in layers III and IV (Giguere and Goldman-Rakic, 1988). However, in rats, MD also projects to cortical layer I (Herkenham, 1980). In contrast, VA and VL projections terminate in both superficial and deep layers (Nakano et al., 1992). VA and VL terminals in the ventrolateral part of BA 4 are primarily located in superficial layers (superficial layer I), while VA and VL terminals in the dorsorostral and ventral part of BA 6, and in BA 8 and 9 are primarily located in deep layers (layers III through VI). VA and VL terminals in the rostral part of BA 4, dorsocaudal part of BA 6, and SMA (part of BA 6) are found in both superficial and deep layers. Terminals of projections originating from the ventral VL are exclusively found in deep layers of rostral BA 4 and dorsocaudal region of BA 6, and those originating from ventral posterior VL are exclusively found in deep layers of BA 4 and 3.

## 7. Reciprocity of thalamocortical connections

Although we have established that projections from MD, VA and VL to some cortical areas are reciprocated, reciprocity might not extend to the neuronal level – the cortical neurons that project to a thalamic nucleus may not receive inputs from the same nucleus. Reciprocity on the neuronal level may either be monosynaptic or multisynaptic. In the monosynaptic case, axons originating in a specific thalamic nucleus synapse onto the cortical neurons that send direct projections back to the same thalamic nucleus. Given that dendrites of cortical pyramidal neurons extend across multiple layers, monosynaptic contacts may be formed by axon terminals from a specific thalamic nucleus that synapse onto dendrites in layers I through IV of the layer V or VI pyramidal neurons that project to thalamus. In the multisynaptic case, axons originating in a specific thalamic nucleus synapse onto a cortical neuron that project to a second neuron in the same cortical area (e.g., in the same cortical microcolumn) that sends projections back to the same thalamic nucleus.

Reciprocity on a neuronal level cannot be established by immunohistochemical studies alone and little is known about reciprocity on a neuronal level in primates. In rodents, in-vitro studies in brain slices combining optogenetics and patch-clamp single cell recordings have shown that both MD and the ventromedial thalamic nucleus form reciprocal connections with the prelimbic cortex on the neuronal level (Collins et al., 2018). MD and the ventromedial thalamus, respectively, induce supra- and subthreshold activity in layer II/III cortical neurons in the prelimbic cortex. These layer II/III neurons, in turn, induce activity in layer V neurons that project to MD and ventromedial thalamus (Collins et al., 2018). Similarly, the medial ventromedial thalamus forms reciprocal connections on the neuronal levels with the anterior lateral motor cortex, and the lateral ventromedial thalamus with the primary motor cortex (Guo et al., 2018). These loops seem segregated, however some studies point to the possibility of crosstalk between them (Yamawaki and Shepherd 2015). Overall, these studies demonstrate that the MD-VA-VL complex and specific parts of the prefrontal and motor cortex form parallel, multi-synaptic reciprocal loops, but that some convergence might exist between these loops.

## 8. Conclusions

In primates, the organization of cortico-thalamo-cortical loops between MD, VA, VL and the cortex is more complex than previously acknowledged. Here, we highlight the key points:

1. Connections between MD, VA, VL and the cortex are less segregated than implied by existing models. MD and VA both have reciprocal connections with the medial, orbital, dorsolateral, and polar prefrontal cortex, as well as the rostral and caudal motor cortex. VL has reciprocal connections with the dorsolateral prefrontal, rostral motor and caudal motor cortex. Furthermore, MD, VA and VL provide input to the posterior parietal cortex, superior parietal lobule, temporal lobe, and posterior cingulate cortex. MD and VA also provide input to the insular cortex. MD projections to superior parietal lobule, temporal lobe, and insular cortex are reciprocated, and VL receives inputs from some parts of the parietal cortex.

2. However, there is a gradient in connectivity patterns between the medial-anterior and the lateral-posterior parts of the MD-VA-VL complex. The medial-anterior parts of the MD-VA-VL complex are more heavily connected with the medial and orbital prefrontal cortex, the central parts of the MD-VA-VL complex are more heavily connected with the dorsolateral and polar prefrontal cortex, and the lateral-posterior parts of the MD-VA-VL complex are more heavily connected with the motor cortex.

3. There is also a connectivity gradient between the medial and lateral subregions of MD. MDmc is more heavily connected with the medial and orbital prefrontal cortex, MDpc is more heavily connected with the dorsolateral and polar prefrontal cortex, and the lateral MD is more heavily connected with the motor cortex.

4. In primates, there is little knowledge about reciprocity of connections between MD, VA, VL and the cortex on the neuronal level. Rodent studies suggest reciprocal connections that form closed loops with possible crosstalk between the loops. Similar studies in primates would be illuminating for understanding how anatomical connections support thalamic response dynamics and function.

5. All parts of MD have connections with some parts of the motor cortex. Considering that these inputs dominantly originate from layer VI, we speculate that input from the motor cortex to MD modulates inputs from the prefrontal cortex. Motor inputs regulate which prefrontal cortical inputs are sustained and enhanced by MD. Likewise, because VL receives inputs from layer VI of some prefrontal regions, we speculate that prefrontal cortical inputs to VL gate incoming motor signals, sustaining or enhancing motor inputs that will result in a correct or advantageous action and suppressing other motor inputs.

More research is required to reach an understanding of how cortico-thalamo-cortical loops between MD, VA, VL and cortex contribute to motor control, action planning, sensorimotor processing and cognition in primates. This review is only a first step in furthering our understanding of this complex network of connections.  We hope that aggregating the current knowledge will highlight where additional work is needed and ultimately aid development of systems-level, computational models to better understand the function of intricate thalamocortical connections.

**Acknowledgements**

This article was supported by the National Institute of Mental Health (R01MH127375), the Simons Collaboration on the Global Brain (Simons foundation grant GB-Culmination-00002986-02), and the Pew Innovation Fund (Pew foundation grant 00037214).